\documentclass[twocolumn,showpacs,floats,floatfix,superscriptaddress,aps,pra]{revtex4-1}
\usepackage{amsfonts}
\usepackage{amssymb}
\usepackage{amsmath}
\usepackage{calc}
\usepackage{graphicx}
\usepackage{bm}

\usepackage[normalem]{ulem}
\usepackage{amsmath,amssymb}
\usepackage{multirow}
\usepackage{xcolor,soul}
\usepackage{srcltx}
\usepackage{hyperref,graphicx}

\def\be{ \begin{equation}}
\def\ee{ \end{equation}}
\def\bea{ \begin{eqnarray}}
\def\eea{ \end{eqnarray}}
\def\bse{ \begin{subequations}}
\def\ese{ \end{subequations}}
\def\bc{ \begin{center}}
\def\ec{ \end{center}}

\begin{document}

\author{Stefano Longhi}
\affiliation{Dipartimento di Fisica, Politecnico di Milano and Istituto di Fotonica e Nanotecnologie del Consiglio Nazionale delle Ricerche, Piazza L. da Vinci 32, I-20133 Milano, Italy}
\title{Nonadiabatic robust excitation transfer assisted by an imaginary gauge field}
\date{\today}

\begin{abstract}
A nonadiabatic and robust method of excitation transfer in a non-Hermitian tight-binding linear chain, assisted by an imaginary gauge field, is theoretically proposed. The gauge field undergoes a linear gradient in time, from a negative to a positive value, which results in an effective transfer of excitation between the two edge sites of the chain. An imaginary (gain/loss) gradient of site energy potentials is introduced to exactly cancel nonadiabatic effects, thus providing an effective shortcut to adiabaticity and pseudo-Hermitian dynamics. Numerical simulations indicate that the non-Hermitian excitation transfer method  is very robust against disorder in hopping rates and site energy of the chain.
\end{abstract}

\maketitle

\section{Introduction}

Coherent transfer of excitations in classical or quantum systems described by effective tight-binding networks is of major interest in different areas of science with a plethora of applications including manipulation of populations in atomic and molecular systems \cite{r1,r2,r3}, control of chemical reactions \cite{r4,r5},
 coherent quantum state transfer and quantum information processing \cite{r6,r7,r8,r9,r10,r11,r12}, efficient transport in organic molecules \cite{r13}, waveguide optics \cite{r14,r15} and atomtronics \cite{r16} to mention a few. Different excitation transfer schemes have been proposed and experimentally demonstrated over the past two decades \cite{r1,r2,r3,r6,r7,r8,r9,r10,r11,r12},  including probabilistic state transfer in a chain with uniform parameters \cite{r6}, perfect state transfer in time-independent chains with
properly tailored hopping amplitudes \cite{r7,r8,r9,r15}, state transfer using externally applied time-dependent control fields \cite{r12}, topologically-protected state transfer protocols \cite{r17,r18}, and state transfer assisted by gauge fields \cite{r19}. Adiabatic protocols, such as those based on the stimulated Raman adiabatic
passage (STIRAP) methods \cite{r1,r3,r11} or topological pumping \cite{r18}, are attractive being rather robust against structural imperfections of the system, however they usually take a long time requiring a slow evolution of the system in one of its adiabatic eigenstate. To realize excitation transfer in a shorter time with a high fidelity, methods of shortcuts to adiabaticity have been proposed and investigated in several studies \cite{r20,r21}. However, these schemes are generally more sensitive to perturbations or disorder in the system than the corresponding adiabatic methods.\par
Excitation transfer methods in open systems, described by effective non-Hermitian Hamiltonians, have been investigated in a few recent works as well \cite{r21,r22,r23}, revealing how
dissipation, gain and dephasing effects can be fruitfully exploited to improve the excitation transfer process and to realize possible routes of shortcut to adiabaticity. In particular, a $\mathcal{PT}$-symmetric extension of the perfect state transfer protocol has been recently proposed in Ref.\cite{r22}, whereas non-Hermitian versions of STIRAP have been suggested in Refs. \cite{r21,r23}. Non-Hermitian extensions of other Hamiltonian models generally studied in quantum state transfer problems and showing quantum phase transitions, such as the isotropic and anisotropic quantum spin models \cite{r23bis}, the Bose-Hubbard models \cite{r23tris}, the Rice-Mele model \cite{r23quatris}, the Kiatev model \cite{Kita}, and the Lipkin-Meshkov-Glick model \cite{Mos} have been suggested as well.\\ 
One of the simplest example of a non-Hermitian tight-binding lattice is provided by the Hatano-Nelson model, which describes the hopping dynamics of a quantum particle on a tight-binding lattice threaded by an imaginary magnetic flux \cite{r24}. 
In their pioneering work, Hatano and Nelson showed that, contrary to an ordinary real magnetic flux leading to a Peierls phase substitution of the hopping rates, an imaginary magnetic field in a disordered one-dimensional lattice can induce a delocalization transition, i.e. it can prevent Anderson localization \cite{r24}. Such a phenomenon, referred to as non-Hermitian delocalization transition, has received great attention in the past two decades \cite{r25,r26,r27,r28}. In particular, unidirectional and bidirectional non-Hermitian transport in the Hatano-Nelson model, which is insensitive to disorder and structural imperfections of the lattice, has been investigated in a few recent works \cite{r26,r27}. While the realization of a synthetic imaginary magnetic field in the solid-state context is challenging, a rather simple optical implementation of the Hatano-Nelson
model, based on photonic transport in coupled optical microrings with tailored gain and loss regions, has been suggested in Refs.\cite{r25,r26}. Such a photonic system has renewed the interest in the Hatano-Nelson model and is expected to provide a viable route toward an experimental observation of the non-Hermitian delocalization transition.\\
 In this article we theoretically propose a nonadiabatic method of robust excitation transfer in a non-Hermitian Hatano-Nelson tight-binding linear chain, which is assisted by an imaginary gauge field. When the gauge field is linearly ramped in time, from a negative to a positive value, any eigenstate of the system evolves localizing the excitation from one edge of the chain, at initial time, to the other edge of the chain at final time. A gain/loss gradient at the chain sites exactly cancels nonadiabatic effects, thus providing an effective shortcut to adiabaticity and fast state transfer. The non-Hermitian transfer method  assisted by the time-varying imaginary gauge field is shown to properly work even when the system is not initially prepared in one of its eigenstate and turns out to be robust against disorder in hopping rates and site energy of the chain.


\section{Nonadiabatic excitation transfer assisted by an imaginary gauge field: theoretical analysis}

Let us consider a linear chain of Wannier states $|n \rangle$ with homogeneous hopping rate $\kappa$ between adjacent sites and threaded by a time-dependent imaginary gauge field $h=h(t)$, as schematically shown in Fig.1(a). For the sake of definiteness, we assume an odd number $(2N+1)$ of sites, however the analysis holds for an even number of sites as well.  Indicating by $ \gamma_n$ the imaginary energy potential at site $| n \rangle$, in the tight-binding approximation and for open boundary conditions the Hamiltonian of the system reads
\begin{eqnarray}
\hat{H}(t) & = & \kappa \sum_{n=-N}^{N-1} \left\{ \exp(-h) | n \rangle \langle n+1 |+ \exp(h) |n+1 \rangle \langle n | \right\}\nonumber \\
& - & i \sum_{n=-N}^N \gamma_n | n \rangle \langle n |. 
\end{eqnarray}
A possible physical realization of a time-dependent imaginary gauge field $h(t)$, which is based on fast modulation of the complex energy site potentials of a lattice, is discussed in the Appendix A.
Note that the Hamiltonian (1) reduces to the standard Hermitian form of a tight-binding chain with uniform hopping rate
\[
\hat{H}_{Herm}= \kappa \sum_{n=-N}^{N-1} \left( | n \rangle \langle n+1 |+ |n+1 \rangle \langle n | \right)
\]
when $h=\gamma_n=0$. Such a simple Hamiltonian is known to realize probabilistic excitation transfer between the two edge sites of the chain at optimal interaction time \cite{r6}. For the chain with uniform hopping amplitudes, the excitation transfer is however not perfect since the energy spectrum of $\hat{H}_{Herm}$, given by the set of energies $E_l$
\begin{equation}
E_l  =  2 \kappa \cos \left[ \frac{\pi l}{2(N+1)} \right]
\end{equation}
($l=1,2,...,2N+1$), is not equally spaced. In fact, to realize perfect excitation transfer from site $n=-N$ to site $n=N$ in a time $2T$, the Hamiltonian should be mirror symmetric and the energy eigenvalues $E_l$ should satisfy the constraint $\exp(-2i E_l T)=(-1)^l \exp(i \alpha)$ for some arbitrary phase  $\alpha$ \cite{r7}. The latter constraint can be clearly satisfied for an equally-spaced energy spectrum, which however requires non-uniform hopping rates \cite{r7,r9}. While the Hamiltonian  (1) with $h=\gamma_n=0$ is mirror symmetric, its energies do not satisfy the constraint given above for any time $2T$, indicating that perfect excitation transfer can not realized.

On the other hand, for $h(t)=h$ constant and $\gamma_n=0$, the Hamiltonian (1) reduces to the non-Hermitian
Hatano-Nelson model without disorder \cite{r24}. In this case, for open boundary conditions $\hat{H}$ is pseudo-Hermitian, the imaginary gauge field does not modify the the energy spectrum of $\hat{H}$, however it provides exponential localization of the eigenstates. For a nonvanishing imaginary gauge field $h$, the eigenstates $|E_l \rangle$ of $\hat{H}$ with $\gamma_n=0$ read explicitly
\begin{equation}
|E_l \rangle  = \frac{1}{\sqrt{N+1}}  \sum_{n=-N}^{N} \exp(hn) \sin \left[ \frac{\pi l (n+N+1)}{2(N+1)} \right]  |n \rangle
\end{equation}
where $l=1,2,...,2N+1$ is the mode index. According to Eq.(3), for $h<0$ the excitation is mainly localized at the left edge $n=-N$ of the chain, whereas for $h>0$ the excitation is mostly localized at the right edge $n=N$. Hence, if the system is initially prepared in one of the eigenstates of $\hat{H}$ with $h<0$ and the imaginary gauge field $h$ is adiabatically increased to a positive value, an effective transfer of the excitation from the left to the right edges of the chain is obtained.

 In order to clarify the transfer method and to remove the adiabaticity constraint, let us assume that the gauge field $h(t)$ is linearly ramped from the negative value $-h_{max}$ at initial time $t_i=-T$ to the positive value $h_{max}$ at final time $t_f=T$, i.e.
\begin{equation}
h(t)=\alpha t
\end{equation}
\begin{figure}[tb]
\includegraphics[width=8 cm]{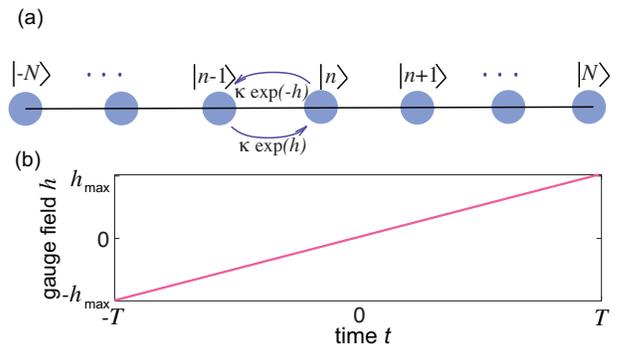}
\caption{(a) Schematic of a linear tight-binding chain, comprising an odd number $(2N+1)$ of sites, with open boundary conditions and with an applied imaginary gauge field $h=h(t)$. (b) The gauge field $h(t)$ is raised from a negative ($-h_{max}$) to a positive ($h_{max}$) value over the interaction time $2T$ with a slope $\alpha=h_{max}/T$.}
\end{figure}

\begin{figure}[tb]
\includegraphics[width=8.5cm]{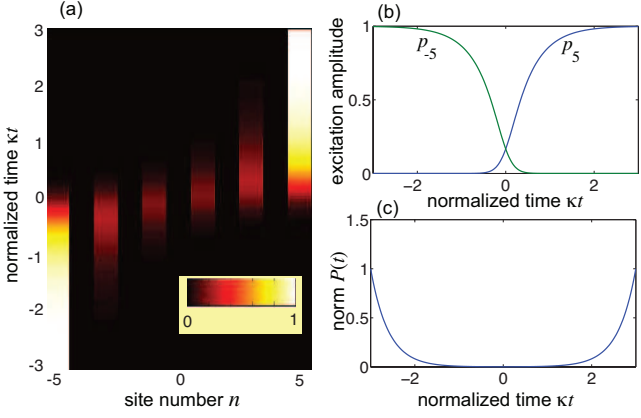}
\caption{Excitation transfer dynamics in a chain composed by $(2N+1)=11$ sites assisted by a time varying imaginary gauge field for parameter values $\kappa T=3$ and $h_{max}=3$.  At initial time $t_i=-T$ the system is prepared in its zero-energy instantaneous eigenstate [Eq.(11) with $l=N+1$]. Imaginary site potentials $\gamma_n=-\alpha n$ are  introduced to exactly cancel non-adiabatic terms. (a) Temporal evolution on a pseudocolor map of the normalized excitation distribution $p_n(t)$. (b) Detailed evolution of the normalized excitations $p_{-5}(t)$ and $p_5(t)$ at the two edge sites of the chain. (c) Temporal behavior of the norm $P(t)=\sum_{n=-N}^{N}|c_n(t)|^2$.}
\end{figure}

\begin{figure}[tb]
\includegraphics[width=8.5cm]{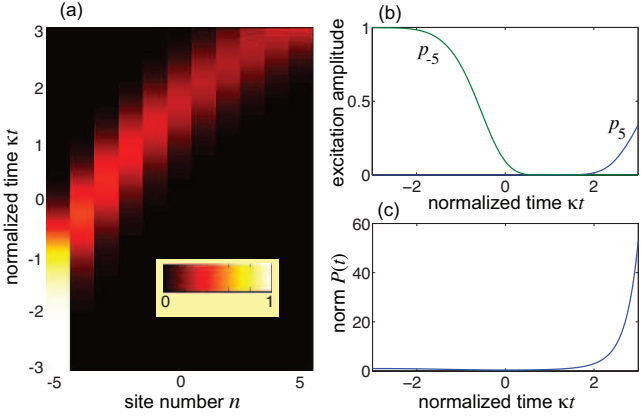}
\caption{Same as Fig.2, but without cancellation of the non-adiabatic terms ($\gamma_n=0$).}
\end{figure}

\begin{figure}[tb]
\includegraphics[width=8.5cm]{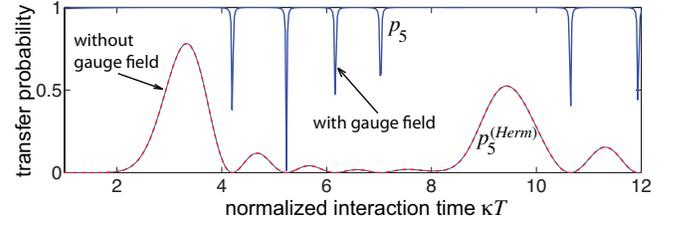}
\caption{Behavior of the transfer probability $p_5$ (solid curve), assisted by the imaginary gauge field, for increasing values of $T$  and for the initial excitation $c_n(t_i)=\delta_{n,-5}$. Parameter values are $(2N+1)=11$, $h_{max}=4$, and $\gamma_n=-\alpha n$. The dashed curve is the transfer probability $p_{5}^{(Herm)}$  in the Hermitian limit, i.e. in the absence of the gauge field ($h=\gamma_n=0$). The behavior of the norm $P$ versus $\kappa T$ is almost overlapped with the dashed curve.}
\end{figure}

\begin{figure}[tb]
\includegraphics[width=8.5cm]{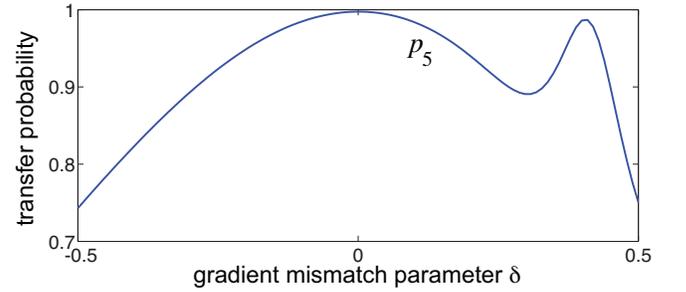}
\caption{Impact of a gradient mismatch on the transfer probability. Behavior of the transfer probability $p_5$, assisted by the imaginary gauge field, in the same chain of Fig.2, comprising $(2N+1)=11$ sites, for $\kappa T=3$, $h_{max}=3$ and $\gamma_n=-(1+\delta) \alpha n$. Perfect cancellation of non adiabatic terms occurs for $\delta=0$. }
\end{figure}

with $\alpha= h_{max}/T$ [Fig.1(b)]. The change of the imaginary gauge field is adiabatic provided that $\alpha \ll \kappa$, whereas non adiabatic effects are expected to arise when the gradient $\alpha$ gets comparable or larger than the hopping rate $\kappa$. However, as we will show below, non adiabatic terms can be exactly cancelled by properly tailoring the imaginary potential site energies $\gamma_n$ in the chain. After expanding the state vector of the system $| \psi(t) \rangle$ on the Wannier basis $|n\rangle$ as $| \psi(t) \rangle = \sum_{n=-N}^{N}c_n(t) | n \rangle$, the evolution equations of the amplitudes $c_n$ read explicitly
\begin{equation}
i \frac{dc_n}{dt}=\kappa \left\{ \exp(-h) c_{n+1}+ \exp(h) c_{n-1} \right\}-i \gamma_n c_n
\end{equation}
with $c_{-(N+1)}=c_{N+1}=0$ for open boundary conditions.
Note that, since the Hamiltonian $\hat{H}(t)$ is not Hermitian, the norm (total probability) defined by the standard inner Dirac product
\begin{equation}
P(t)=| \langle \psi(t) | \psi(t) \rangle |^2=\sum_{n=-N}^N |c_n(t)|^2
\end{equation}
is not conserved in the dynamics. This feature is common to other non-Hermitian extensions of excitation transfer methods, such as the $\mathcal{PT}$-symmetric extension of the perfect state transfer model previously introduced in Ref.\cite{r22}.  To quantify the goodness of the transfer method for a non-conserving norm, we consider the normalized distribution $p_n(t)$ of the excitation at site $|n \rangle$, defined as
 \begin{equation}
p_n(t)=\frac{|c_n(t)|^2}{P(t)}=\frac{|c_n(t)|^2}{\sum_{n=-N}^{N}|c_n(t)|^2}.
 \end{equation} 
Let us now introduce the imaginary gauge transformation
 \begin{equation}
 c_n(t)=a_n(t) \exp[h(t) n]
 \end{equation}
 so that the evolution equations of the amplitudes $a_n(t)$ read
 \begin{equation}
i \frac{da_n}{dt}=\kappa ( a_{n+1}+ a_{n-1}) -i \gamma_n a_n-i n \alpha a_n
\end{equation}
where the last term of the right hand side of Eq.(9) accounts for nonadiabatic effects in the dynamics, i.e. a non-negligible gradient $\alpha$ of the gauge field. 
 Interestingly, provided that the imaginary site potential energies $\gamma_n$ are tailored to satisfy the condition
 \begin{equation}
 \gamma_n=-n \alpha
 \end{equation}
 the nonadiabatic terms in Eq.(9) are exactly cancelled, and the system evolves remaining in its instantaneous eigenstate. Precisely, if at initial time $t_i=-T$ the system is prepared in one of its instantaneous eigenstate
 \begin{equation}
 c_n(t_i)= \exp\left[-h_{max}(n+N) \right] \sin \left[ \frac{\pi l (n+N+1)}{2(N+1)} \right] 
 \end{equation}
 for some index $l=1,2,...,2N+1$, corresponding to localization of the excitation at the left edge $n=-N$ of the chain, the final state of the system at time $t_f=T$ is {\it exactly }given by 
 \begin{eqnarray}
 c_n(t_f) & = & \exp \left[ h_{max}(n-N) \right] \sin \left[ \frac{\pi l (n+N+1)}{2(N+1)} \right]  \nonumber \\
  & \times & \exp(-2i  E_l T), 
 \end{eqnarray}
 corresponding to localization of the excitation at the right edge $n=N$ of the chain (see Appendix B for technical details). More generally, provided that the nonadiabatic terms are exactly cancelled, Eq.(9) indicates that the time-dependent Hamiltonian $\hat{H}(t)$ defined by Eq.(1) is pseudo-Hermitian, i.e. its evolution can be obtained from the time-independent Hermitian chain (9) after the imaginary gauge transformation (8). As an example, Fig.2 shows a typical temporal evolution of the normalized distribution $p_n(t)$ of the excitation at site $|n \rangle$ in a linear chain comprising $(2N+1)=11$ sites with the system initially prepared in the zero-energy eigenstate $l=N+1$, i.e. $c_n(t_i)=\pm \exp[-h_{max}(n+N)]$ for $n=-N,-N+2,-N+4,..., N-2,N$ and $c_n(t_i)=0$ otherwise. Note that for three sites $2N+1=3$ the transfer method can be regarded as a non-Hermitian extension of the STIRAP technique in the Hermitian case, where the system adiabatically evolves remaining in its dark state (the middle site is never populated). As compared to STIRAP, where the hopping rates are changed in time, in our method shortcut to adiabaticity is much simpler since it just requires to introduce an imaginary linear gradient of site potential energies [Eq.(10)]. In Fig.2 complex site energy potentials $\gamma_n=-n \alpha$ are introduced to exactly cancel non-adiabatic terms in the dynamics. Note that an effective transfer of excitation form the left to the right edge sites of the chain is obtained, with the norm which is conserved at the end of the interaction in spite of non-Hermitian dynamics. The behavior of the norm, shown in Fig.2(c), can be physically explained as follows. Let us first consider a slow (adiabatic) change of the gauge field. In the first time interval $(t_i=-T,0)$, the imaginary gauge field $h(t)=\alpha t$ is negative ($h<0$) and a forward-propagating wave experiences a power attenuation owing to the dispersion relation  of the Hatano-Nelson lattice \cite{r26,r27}: therefore, in the first stage of the transfer the norm decreases as a result of dissipation of a forward-propagating wave. Conversely, in the second stage of the transfer, i.e. in the time interval $(0,t_f=T)$, the gauge field is positive ($h>0$) and a forward-propagating wave is now amplified (rather than attenuated) in the lattice because of flipping of the imaginary part of the lattice energy band \cite{r26,r27}. Wave amplification in the second stage of excitation transfer exactly compensates for wave attenuation in the first stage of transfer, thus resulting in the conservation of the norm at the final time $t_f=T$. For a rapid (non-adiabatic) change of the gauge field a gradient of site potential energies, i.e. loss/gain terms $\gamma_n=-n \alpha$, are also  responsible for non-unitary dynamics: in the first stage  the wave is also damped because dissipation occurs in the lossy sites $n <0$ of the chain while the excitation is being transferred from the left edge site $n=-N$ toward the center of the chain. Conversely, in the second time interval $(0, t_f=T)$ the norm increases because the excitation is now amplified in the gain sites $n>0$ of the chain, till reaching the right edge site with conserved norm.
 Figure 3 shows, for comparison, the evolution of $p_n(t)$ when $\gamma_n=0$, i.e. without the nonadiabatic correction terms. Note that in this case degradation of the state transfer is clearly observed.
 
 The previous analysis requires, strictly speaking, that the initial excitation of the system at time $t=t_i$ is one of the $(2N+1)$ eigenstates of $\hat{H}(t_i)$, which shows strong (exponential) localization on the left edge site of the chain with a degree of localization that increases as the imaginary gauge field $h_{max}$ is increased [see Eq.(11)]. However, it is of major importance to check whether the transfer method holds even when the initial excitation deviates from one of the eigenstates, for example in the most common case where at initial time the chain is excited in the left edge site solely, i.e. for the initial condition $c_n(t_i)=\delta_{n,-N}$. In this case, provided that the non-adiabatic terms in the dynamics are exactly cancelled [Eq.(10)], it can be readily shown that at final time $t_f$ the excitation amplitudes of the various sites in the chain are given by (see Appendix B)
 \begin{equation}
 c_n(t_f)=\exp [h_{max}(n-N)]  \theta_n
 \end{equation}
 where 
 \begin{eqnarray}
 \theta_n & \equiv & \frac{1}{N+1} \sum_{l=1}^{2N+1} \sin \left[ \frac{\pi l (n+N+1)}{2(N+1)} \right] \nonumber \\
 & \times & \sin \left[   \frac{\pi l}{2(N+1)} \right] \exp ( -2i T E_l). 
\end{eqnarray}
is the distribution of excitation in the chain at final time $t_f=T$ that one would obtain in the Hermitian limit $h=\gamma_n=0$. Note that $p_{N}^{(Herm)}=|\theta_N|^2$ is the transfer probability that one would observe in the Hermitian chain. Provided that $\theta_N$ is sufficiently far from zero, Eq.(13) shows that at the final time $t_f$ the excitation is again mostly localized at the right edge site $n=N$ of the chain with a transfer probability given by $p_N$, i.e. the transfer method works properly also when the initial state at time $t_i$ is not exactly one eigenstate of $\hat{H}(t_i)$. However, the norm of the final state, $P(t_f)$, is diminished as compared to the initial value $P(t_i)=1$, indicating that the excitation transfer is a dissipative process. In particular, for $h_{\max}$ larger than $\sim 1$, an estimate of the final norm is given by $P(t_f) \simeq |\theta_N|^2=p_{N}^{(Herm)}$, as shown in Appendix B. Hence, to minimize the loss of the norm $P(t_f)$, the interaction time $2T$ should be properly chosen to optimize $p_N^{(Herm)}$. 
Figure 4 shows, as an example, the behavior of the transfer probability $p_N=|c_N(t_f)|^2/P(t_f)$ for increasing values of the normalized  interaction time $\kappa T$ and for $N=5$. In the figure the behavior of the transfer probability in the uniform Hermitian chain, $p_{N}^{(Herm)}$, is also shown for comparison. Note that for almost any interaction time $T$ one has $p_N \simeq 1$ even though $p_{N}^{(Herm)}$ is considerably smaller than one, indicating that the gauge field greatly improves the fidelity of transfer as compared to the static Hermitian chain. Only at some discrete values of $\kappa T$, where $p_{N}^{(Herm)}$ vanishes, the transfer probability $p_N$ may deviate form one and  the gauge-assisted transfer method fails.\\
Finally, it is worth mentioning that cancellation of the non adiabatic terms, which ensures efficient excitation transfer between the edge sites in the chain,   requires that relations (4) and (10) be simultaneously satisfied. i.e. that the gradient of the imaginary (loss/gain)  site potential energies, $\gamma_n=-\alpha n$, be equal to the rate of increase of the imaginary gauge field, $h(t)=\alpha t$. In practice, however, some deviations of the gradients are expected to arise. To check the sensitivity of the excitation transfer method versus a gradient mismatch, we considered the case of imperfect cancellation of non adiabatic terms by replacing Eq.(10) with the more general relation $\gamma_n=-(1+\delta) \alpha n$, where  the dimensionless parameter $\delta$ measures the mismatch from the ideal case $\delta=0$. Figure 5 shows, as an example, the behavior of the normalized transfer probability $p_5$ versus the mismatch parameter $\delta$ in the same linear chain of Fig.2, comprising $(2N+1)=11$ sites, for parameter values $\kappa T=3$ and $h_{max}=3$. Note that a high transfer efficiency, larger than $ \sim 90 \%$, is observed for $|\delta| <0.2$, i.e. for quite large (up to $ \sim 20 \%$) gradient mismatch from the ideal condition.    

\section{Excitation transfer in a linear chain with disorder}

\begin{figure*}[tb]
\includegraphics[width=16cm]{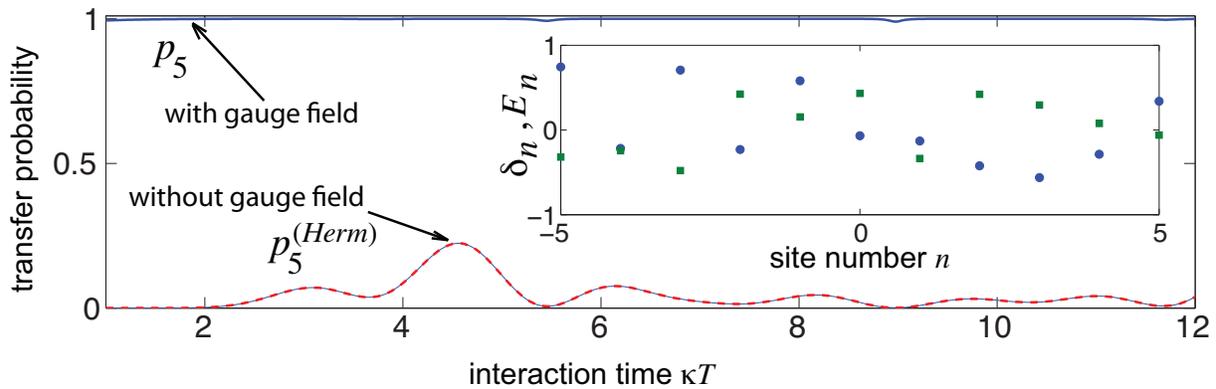}
\caption{Same as Fig.4, but for a chain with disorder in hopping rates $\delta_n$ and site energies $E_n$. The dashed curve shows the transfer probability $p_{5}^{(Herm)}$ versus interaction time $T$ in the Hermitian chain ($h=\gamma_n=0$), whereas the solid curve depicts the transfer probability $p_5$ versus $T$ with the imaginary gauge field. The distribution of disorder $\delta_n$ (squares) and $E_n$ (dots) is shown in the inset. Other parameter values are as in Fig.4.}
\end{figure*}

\begin{figure*}[tb]
\includegraphics[width=16cm]{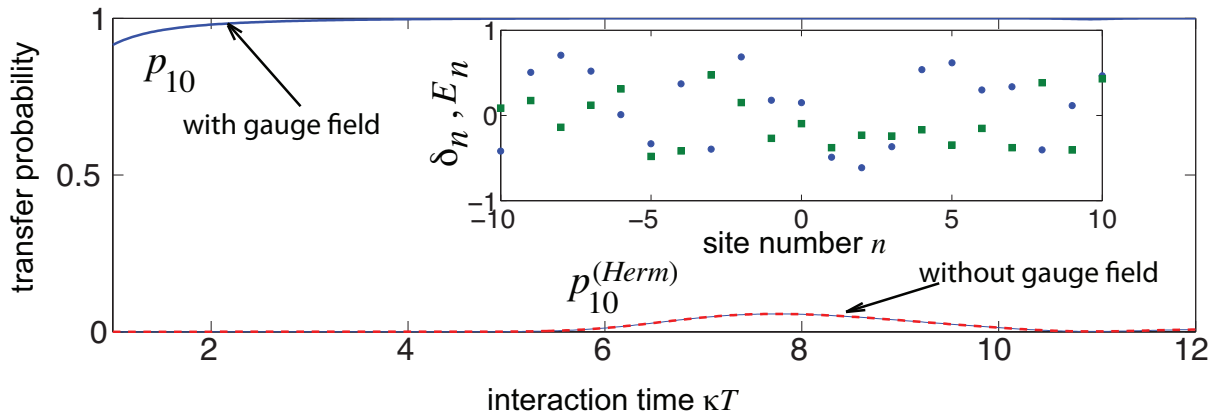}
\caption{Same as Fig.6, but for a chain with $(2N+1)=21$ sites.}
\end{figure*}

An interesting feature of the excitation transfer method assisted by an imaginary gauge field is its robustness against lattice imperfections and disorder. Let us consider a linear chain comprising $(2N+1)$ sites with disorder in either or both site energy and hopping rates described by the Hermitian Hamiltonian
\begin{eqnarray}
\hat{H}_{Herm} & = & \sum_{n=-N}^{N-1} \kappa(1+\delta_n)  \left( | n \rangle \langle n+1 |+ |n+1 \rangle \langle n | \right) \nonumber \\
& + &  \sum_{n=-N}^{N} E_n |n \rangle \langle n | 
\end{eqnarray}
where $\delta_n$ and $E_n$ are random variables that account for disorder in the hopping rate and site energies. Owing to disorder, the eigenstates of $\hat{H}_{Herm}$ become localized and probabilistic excitation transfer in a long chain is degraded, as shown as an example in Figs.6 and 7 for a chain comprising $(2N+1)=11$ sites and $(2N+1)=21$ sites, respectively. The dashed curves in the figures show the numerically-computed transfer probability versus normalized interaction time as in Fig.4, but in the presence of disorder. In the figures, the distributions of  $\delta_n$ and $E_n$, shown in the insets,   are two realizations of disorder as obtained by assuming $\delta_n$ and $E_n$ random variables with uniform distributions in the range $(-1,1)$. The application of the imaginary gauge field $h$ modifies the localization properties of eigenstates and can thus prevent Anderson localization and restore transport along the chain \cite{r24,r25,r26}. Therefore, we expect that the non adiabatic transfer method introduced in the previous section, based on a linearly-ramped imaginary gauge field, is robust against disorder or structural imperfections of the linear chain. The Hamiltonian of the system, with the imaginary gauge field and imaginary site potential gradient aimed at canceling nonadiabatic terms, read
\begin{eqnarray}
\hat{H} & = & \sum_{n=-N}^{N-1} \kappa(1+\delta_n)  \left\{ \exp(-h) | n \rangle \langle n+1 |+ \right. \nonumber \\
& + & \left.  \exp(h) |n+1 \rangle \langle n | \right\} \nonumber \\
& + &  \sum_{n=-N}^{N} (E_n -i \gamma_n) |n \rangle \langle n | 
\end{eqnarray}
where $h(t)=\alpha t$, $\gamma_n=-n \alpha$, $\alpha= h_{max}/T$, and $2T$ is the interaction time. Let us expand the state vector of the system $|\psi(t) \rangle$ on the Wannier basis by letting $|n \rangle$, $|\psi(t) \rangle=\sum_{n=-N}^{N} c_n(t) | n \rangle$. The evolution equations of the amplitudes $c_n(t)$ read
\begin{eqnarray}
i \frac{dc_n}{dt} & = & \kappa(1+\delta_n) \exp(-h) c_{n+1}+ \kappa(1+\delta_{n-1}) \exp(h) c_{n-1} \nonumber \\
& + & (E_n-i \gamma_n)c_n
\end{eqnarray}
which differs from Eq.(5) because of the disorder $\delta_n$ and $E_n$ in hopping rates and site energies. Let us assume that at initial time $t=t_i=-T$ the chain is excited in its left edge site, i.e. $c_n(t_i)=\delta_{n,-N}$. At final time $t_f=T$, after introduction of the gauge transformation (8) it can be readily shown that the amplitudes $c_n(t_f)$ are given by Eq.(13), where $\theta_n$ is the solution that one would obtain for $h=\gamma_n=0$, i.e. for the disordered Hermitian chain with Hamiltonian (15).  The exponential term on the right hand side of Eq.(13) can overcome Anderson localization, thus resulting in an efficient localization of the excitation at the right edge site $n=N$ and a high transfer probability $p_N$, as shown in Figs.6 and 7 (solid curves). Interestingly, while the disorder degrades the transfer probability $p_{N}^{(Herm)} =|\theta_N|^2$ in the Hermitian case owing to Anderson localization, it prevents the amplitude $\theta_N$ to vanish at almost any interaction time $T$, thus avoiding the dips in the transfer probability $p_N$ in the disordered chain when the gauge field is switched on (compare the solid curve of Fig.4 with those of Figs.6 and 7). In other words, while disorder greatly degrades the transfer probability in the Hermitian chain, it improves the transfer in the non-Hermitian case preventing the failure of the transfer method at certain discrete values of interaction time  $T$.\\
The benefit of the imaginary gauge field in realizing a reliable and disorder-insensitive excitation transfer between the two edge sites of the chain is clearly demonstrated when considering the distribution of the transfer probability for a given interaction time $T$ and for different realizations of disorder. As an example, Fig.8 shows the numerically-computed distribution of the transfer probabilities in a chain comprising $(2N+1)=11$ sites for 10000 realizations of disorder in hopping rates and for a fixed interaction time $T=3.33/ \kappa$, which maximizes the transfer probability in the Hermitian chain without disorder ($p_{N}^{(Herm)}=|\theta_N|^2 \simeq 0.78$). The distribution $p_{N}^{(Herm)}=|\theta_N|^2$ refers to the transfer probability in the Hermitian chain, whereas the distribution $p_N$ refers to the non-Hermitian chain with an applied gauge field $h_{max}=2$. Two different statistics of disorder, namely uniform and normal distributions, have been assumed. Note that, while in the Hermitian chain the transfer probability $p_{N}^{(Herm)}$ is rather sensitive to the realization of disorder and is typically lowered as compared to the ordered chain, in the non-Hermitian chain with the imaginary gauge field the transfer probability $p_N$ is insensitive to disorder and close to $100 \%$ for both uniform and normal distributions. In the latter case the distribution of transfer probability is slightly broadened because of the larger standard deviation $\sigma$ of disorder for the normal distribution ($\sigma=0.5$) as compared to the uniform distribution ($\sigma=1/ \sqrt{12}$).

\begin{figure*}[tb]
\includegraphics[width=16cm]{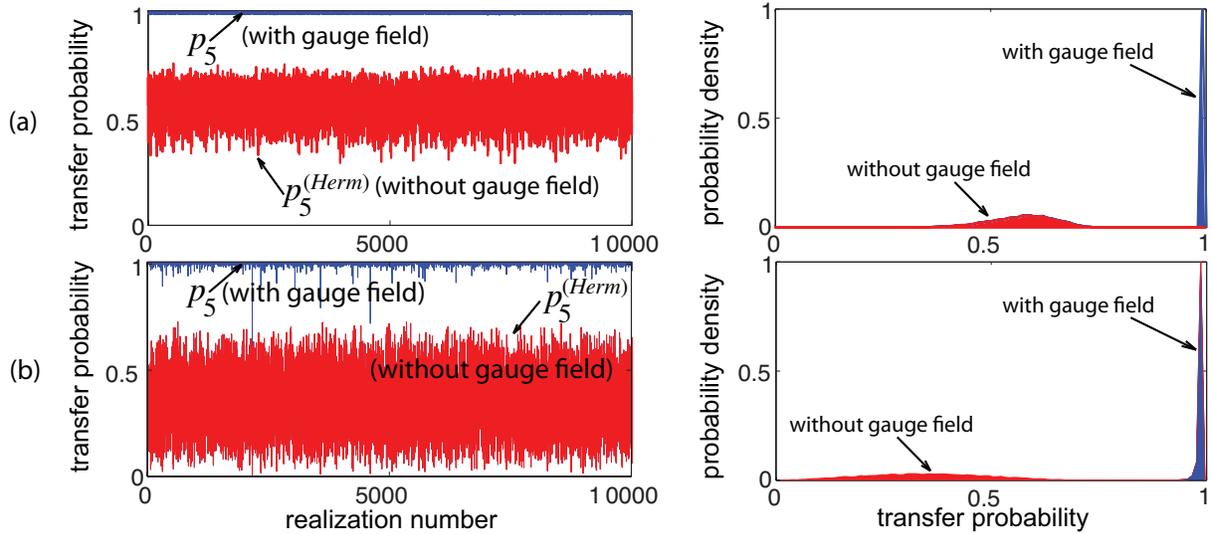}
\caption{Numerically-computed transfer probability in a chain comprising $(2N+1)=11$ sites for 10000 realizations of disorder in the hopping rates and for a fixed interaction time $T= 3.33 / \kappa$. In (a) $\delta_n$ is assumed to be a random variable with uniform distribution in the interval $(-0.5,0.5)$, whereas in (b) $\delta_n$ is assumed to be a random variable with a normal distribution with standard deviation 0.5. The right panels show the statistics of the transfer probability with and without the imaginary gauge field ($h_{max}=2$).}
\end{figure*}

\section{Conclusion and discussion}

Non-Hermitian extensions of Hamiltonian models of major relevance in problems of quantum control, quantum state transfer, quantum or classical transport, and quantum annealing have received a great deal of attention in the past few years \cite{r21,r22,r23,r23bis,r23quatris,Kita, annealing,Mos}. Interestingly, dissipation, gain and dephasing effects can be fruitfully exploited to improve the state transfer process \cite{r23}, to realize possible routes of shortcut to adiabaticity \cite{r21}, to optimize quantum annealing methods \cite{annealing}, and to amplify the entanglement and spin squeezing near quantum phase transitions \cite{Mos}. In this work a fast nonadiabatic method of excitation transfer in a non-Hermitian network, which is robust against structural imperfections and disorder, has been theoretically proposed. The non-Hermitian model under investigation is an extension of the Hatano-Nelson Hamiltonian \cite{r24}, which is known to show a non-Hermitian delocalization transition. Robust transfer is assisted by an imaginary gauge field,  which is linearly increased in time  from a negative to a positive value, resulting in an effective and disorder-insensitive transfer of excitation between the two edge sites of the network. Nonadiabatic effects are exactly cancelled by the introduction of an imaginary gradient of site energy potentials, providing an effective shortcut to adiabaticity and pseudo-Hermitian dynamics. A possible physical implementation of the non-Hermitian model could be provided by light transport in a chain of coupled resonator optical waveguides, where a synthetic imaginary gauge field can be realized in principle by means of auxiliary out-of-resonance cavities with optical gain and loss \cite{r26}, or using the modulation scheme discussed in Appendix A. Our results disclose interesting aspects of classical or quantum transport in non-Hermitian Hamiltonian models and reveal an entirely new platform upon which robust state transfer can be understood and realized. They also may suggest several new directions of research. For example, the application of a space-dependent imaginary gauge field (the imaginary gauge phase $h$ entering in Eq.(1) depends on lattice site), together with appropriate nonadiabatic correction terms, could be used to steer an initial delocalized state into a desired final state in an arbitrarily short time. The interplay of imaginary and ordinary gauge fields in assisting wave transport could be investigated as well, especially in two-dimensional networks where topological protection could come into play. There are also some open questions, for example how to implement imaginary gauge fields in solid-state or matter wave systems and the extension of non-Hermitian transfer models in second quantization framework or for the description of mixed state dynamics \cite{ropen}. 

\appendix

\section{Physical realization of a time-dependent imaginary gauge field}
In this Appendix we briefly discuss a possible physical realization of a time-dependent imaginary gauge field which could be implemented in coupled-resonator optical waveguide (CROW) structures \cite{referee1}, such as  photonic-crystal defect cavities, microspheres, microdisks, and microring resonators. The present scheme, however, is a general one and could be potentially applied to quite arbitrary non-Hermitian lattice systems with modulated complex on-site potential energies. \\
 Let us consider a CROW structure comprising  $(2N+1)$ cavities, each supporting a single mode with amplitude $b_n$ and resonance frequency $\omega_0$. We assume that a linear and time-dependent gradient of the {\em complex} frequencies of the cavities is superimposed to the chain, so that coupled-mode equations describing photon hopping in the chain read (see, for instance, \cite{referee2})
\begin{equation}
i \frac{db_n}{dt}= \rho(b_{n+1}+b_{n-1})+ \omega_0 b_n+n \delta \omega_0(t) b_n
\end{equation}
 ($-N \leq n \leq N$), where $\rho$ is the coupling constant between adjacent cavities. The real part of $\delta \omega_0(t)$ describes the offset rate of the resonance frequency of the dynamically-tuned cavity from the central frequency $\omega_0$, whereas the imaginary part of $\delta \omega_0(t)$ describes a gain/loss term gradient, namely a loss term for $n {\rm Im} (\delta \omega_0(t))<0$ and a gain term for  $n \rm{Im} ( \delta  \omega_0(t))>0$. Ultrafast dynamic modulation of the refraction index, leading to a modulation of the resonance frequency, can be achieved by carrier injection \cite{referee3}, whereas modulation of the gain/loss requires active resonators with modulation of the electrical and/or optical pumping. In writing Eq.(A1) we assumed that a single mode of each cavity in the chain is excited, as it is usual in coupled-mode theory of CROW structures \cite{referee1,referee2}. In case of CROW realized by defects in a photonic crystal, the single mode assumption is justified because of the defect sustains a single resonance, whereas in other CROW structures, such as those based on coupled microring resonators,  single mode operation is generally ensured by the excitation of the chain at one edge with an external coherent field using a bus waveguide \cite{referee1}, which excites a single traveling-wave wishpering gallery mode of the ring.\\
 To realize the effective Hamiltonian given by Eq.(1) in the text, let us assume
 \begin{eqnarray}
 \delta \omega_0(t) & = & i \alpha+\Gamma \cos (\Omega t+ i \phi) \\ 
 & = &  i \alpha +\Gamma \cosh \phi \cos (\Omega t)-i \Gamma \sinh \phi \sin (\Omega t) \nonumber
 \end{eqnarray}
 where $\alpha$,  $\Gamma$, $\Omega$ and $\phi$ are real numbers. Note that the dc term of $\delta \omega_0(t)$ describes a linear gradient of gain/loss in the resonators, whereas the ac term corresponds to a mixed and quarter-phase-shifted sinusoidal modulations at frequency $\Omega$  of the real and imaginary parts of the refractive mode index. The phase $\phi$ is allowed to vary on a slow time  scale as compared to the period $ 2 \pi / \Omega$ of the carrier. After setting
 \begin{equation}
 b_n(t)= \exp[-i \omega_0 t -i \Phi(t) n] c_n(t)
 \end{equation}
with 
\begin{eqnarray}
\Phi(t) & = & \Gamma \int_0^t d \xi \cos(\Omega \xi+ i \phi) \nonumber \\
& = & \frac{\Gamma}{\Omega} \left[ \sin (\Omega t+i \phi)-i \sinh \phi \right]
\end{eqnarray}
substitution of Ansatz (A3) into Eq.(A1) yields the following coupled-mode equations for the amplitudes $c_n$
\begin{equation}
i \frac{dc_n}{dt}=\rho \left[ \exp(-i \Phi) c_{n+1}+ \exp(i \Phi) c_{n-1}) \right]+i n \alpha c_n .
\end{equation}
For an oscillation frequency $\Omega$ much larger than the coupling constant $\rho$, in the rotating-wave approximation we can average the rapidly-oscillating terms on the right-hand side of Eq.(A5). Taking into account that
\begin{equation}
 \langle \exp[\mp i \Phi(t)] \rangle= J_0 \left( \frac{\Gamma}{\Omega} \right) \exp( \mp h)
\end{equation}
 where 
 \begin{equation}
 h \equiv \frac{\Gamma}{\Omega} \sinh \phi,
 \end{equation}
$J_0$ is the Bessel function of first kind and zero order, and $\langle ... \rangle$ denotes the time average over the oscillation period $ 2 \pi / \Omega$, Eq.(A5) finally reads
\begin{equation}
i \frac{dc_n}{dt}=  \kappa \left[ \exp(-h) c_{n+1}+\exp(h) c_{n-1} \right]+ i n \alpha c_n
\end{equation}
where we have set 
\begin{equation}
\kappa= \rho J_0(\Gamma / \Omega).
\end{equation} 
In their present form, Eq.(A8) is thus equivalent to Eq.(5) given in the text with $\gamma_n= -\alpha n$, and thus the CROW structure with a modulated index gradient effectively describes the non-Hermitian Hamiltonian (1) with an imaginary gauge field $h$ defined by Eq.(A7). To realize a synthetic time-varying imaginary gauge field $h(t)=\alpha t$ in the CROW, one should modulate the phase $\phi(t)$ according $ \phi(t)= {\rm asinh} ( \Omega \alpha t / \Gamma)$ while keeping the amplitude $\Gamma$ independent of time.

 \section{Temporal evolution of excitation amplitudes: some analytical results}
In this Appendix we provide some analytical results regarding the temporal evolution of the excitation amplitudes $c_n(t)$ along the chain for the non-Hermitian Hamiltonian (1) with a linearly-varying imaginary gauge field $h(t)=\alpha t$ and with the non-adiabatic correction terms $\gamma_n=-\alpha n$. Owing to Eq.(8), the excitation amplitudes $c_n(t_f)$ at final time $t_f=T$ are related to the amplitudes $c_n(t_i)$ at initial time $t_i=-T$ by the relation
\begin{eqnarray}
c_n(t_f) & = & a_n(t_f) \exp [h(t_f) n]  =   \sum_{l=-N}^N \mathcal{U}_{n,l} a_l(t_i) \exp[h(t_f) n] \nonumber \\
& = &  \sum_{l=-N}^N \mathcal{U}_{n,l} \exp [h(t_f)n-h(t_i)l] c_l(t_i)  \\
& = & \sum_{l=-N}^N \mathcal{U}_{n,l} \exp [h_{max}(n+l)] c_l(t_i) \nonumber
\end{eqnarray}
 where $h_{max}= \alpha T$ and $\mathcal{U}$ is the propagator of Eq.(9) from $t=t_i$ to $t=t_f$, i.e.
 \begin{equation}
 a_n(t_f)= \sum_{l=-N}^{N} \mathcal{U}_{n,l} a_l(t_i).
 \end{equation}
 Note that, provided that the non-adiabatic terms are cancelled by assuming $\gamma_n=-n \alpha$, $\mathcal{U}$ is the propagator of a linear Hermitian chain with uniform hopping amplitude $\kappa$ comprising $(2N+1)$ sites, which is described by a unitary matrix. Its expression is readily constructed from the eigenvectors $| E_l \rangle$ [Eq.(3) with $h=0$] and corresponding eigenvalues $E_l$ [Eq.(2)] of $\hat{H}_{Herm}$, and reads explicitly 
 \begin{eqnarray}
 \mathcal{U}_{n,l} & = & \sum_{\sigma=1}^{2N+1} \langle E_{\sigma} | l  \rangle \langle n | E_{\sigma} \rangle \exp(-2i T E_{\sigma}) \\
 & = & \frac{1}{N+1}\sum_{\sigma=1}^{2N+1} \sin \left[ \frac{ \pi \sigma (l+N+1)}{2(N+1)}\right]  \sin \left[ \frac{ \pi \sigma (n+N+1)}{2(N+1)}\right] \nonumber \\
 & \times &  \exp(-2i T E_{\sigma}). \nonumber
 \end{eqnarray}
 Note that, using Eq.(B1), the norm of the final state, $P(t_f)=\sum_{n=-N}^{N} |c_n(t_f)|^2$, reads
 \begin{equation}
 P(t_f)=\sum_{l, \sigma=-N}^{N} W_{l, \sigma} c_l (t_i) c_{\sigma}^{*}(t_i)
 \end{equation}
 with $P(t_i)=\sum_{n=-N}^{N} |c_n(t_i)|^2=1$ and where we have set
 \begin{equation}
 W_{l,\sigma} \equiv \sum_{n=-N}^{N} \mathcal{U}_{n,l} \mathcal{U}_{n, \sigma}^{*} \exp[h_{max}(2n+l+\sigma)]. 
 \end{equation}
 In the ordinary Hermitian problem, i.e. without the imaginary gauge field $h_{max}=0$, $W_{l, \sigma}= \delta_{l, \sigma}$ is the identity matrix since $\mathcal{U}$ is a unitary matrix, so that the norm is conserved $P(t_f)=P(t_i)=1$. However, in the non-Hermitian case $h_{max} \neq 0$, $W_{l, \sigma}$ deviates from the identity matrix and thus the final norm is generally different than the initial one as a signature of non-Hermitian dynamics.\\
 As a first example, let us assume that at initial time $t_i$ $c_n(t_i)$ is the instantaneous eigenstate of $\hat{H}(t_i)$ with energy $E_l$; see Eq.(11) in the main text. Then, since $a_n(t_f)=a_n(t_i) \exp(-2iTE_l)$, one readily obtains 
 \begin{eqnarray}
 c_n(t_f) & = & a_n(t_t) \exp[h(t_f)n]=a_n(t_i) \exp[h(t_f)n-2iTE_l] \nonumber \\
 & = & c_n(t_i) \exp[h(t_f)n-h(t_i)n -2iTE_l].
\end{eqnarray} 
Taking into account that $h(t_f)-h(t_i)=2 \alpha T=2 h_{max}$, one obtains $c_n(t_f)=c_n(t_i) \exp (2 h_{max} n -2 i T E_l)$, which using Eq.(11) finally yields Eq.(12) given in the main text. In this case, since the system evolves in one of its eigenstates, one can readily shown that $P(t_f)=P(t_i)=1$, i.e. the norm is conserved after the transfer of excitation.\\
As a second example, let us assume that the chain is initially excited in the left hand edge site, i.e. $c_l(t_i)=\delta_{l,-N}$. From Eq.(B1) one obtains 
\begin{equation}
c_n(t_f)= \theta_n \exp[h_{max}(n-N)] 
\end{equation}
where we have set $\theta_n \equiv \mathcal{U}_{n,-N}$. Taking into account the form of $\mathcal{U}_{n,-N}$ given by Eq.(B3), one finally obtains Eqs.(13) and (14) given in the text. In this case the norm of the final state, as obtained from Eqs.(B4) and (B5) with $c_{n}(t_i)=\delta_{n,-N}$, reads
\begin{equation}
P(t_f)=W_{-N,-N}= \sum_{n=-N}^N | \theta_n|^2 \exp[2h_{max}(n-N)]
\end{equation}
where $\theta_n=\mathcal{U}_{n,-N}$.  Note that, since $\sum_{n=-N}^{N} |\theta_n|^2=1$ and $\exp [ 2 h_{max}(n-N)]<1$ for $n < N$, one has $P(t_f) < \sum_{n=-N}^{N} |\theta_n|^2=1$, i.e. the final norm in this case is always smaller than the initial one, indicating that excitation transfer is dissipative. For a sufficiently large value of the gauge field $h_{max}$, provided that $\theta_{N}$ does not vanish from Eq.(B8) it follows that the dominant term in the sum is the one with index $n=N$, so that an estimate of the final norm is given by $P(t_f) \sim |\theta_{N}|^2$.

\end{document}